\documentclass[a4paper,11pt]{article}
\usepackage{jheppub} 
\usepackage{subcaption}
\usepackage{mathrsfs}
\usepackage{dsfont}

\usepackage{physics}
\usepackage{color,xcolor}
\usepackage{bm}
\usepackage{mdframed}

\usepackage{soul} 
\usepackage{amsthm}

\title{\boldmath Pseudo-Nambu-Goldstone bosons and tachyons in flux compactification}

\author[a]{Victor Franken}

\affiliation[a]{CPHT, CNRS, École polytechnique, Institut Polytechnique de Paris\\91120 Palaiseau, France}

\emailAdd{victor.franken@polytechnique.edu}

\abstract{In a six-dimensional gauge theory compactified on a torus with magnetic flux, translational symmetry in the extra dimensions is broken. As a result, a massless Nambu-Goldstone boson appears in the four-dimensional effective Lagrangian. We show that a model with two $U(1)$ gauge symmetries includes a pseudo-Nambu-Goldstone scalar boson, whose mass is finite but shielded from large quantum corrections. This opens the door to achieving scalar masses in the TeV range. Additionally, we explore the presence of tachyons in the effective Lagrangian for $SU(2)$ gauge theories. By introducing a scalar field in the adjoint representation of $SU(2)$ with a finite vacuum expectation value, we demonstrate that tachyonic modes can be eliminated for sufficiently large values of the coupling constant, while preserving fermion chirality in the model.}

\begin{document}

\begin{flushright}

\end{flushright}

\maketitle

\flushbottom
\section{Introduction}
\label{sec:intro}

The Standard Model of fundamental interactions has proven to be a remarkably successful theory, withstanding decades of experimental testing. However, it still leaves some fundamental questions unresolved, such as the hierarchy problem, which has been a central focus in the search for physics beyond the Standard Model. These unresolved issues suggest the need for physics beyond the Standard Model. At higher energy scales, we would expect to find evidence of new physics — deviations from the Standard Model’s predictions. Since no new physics has been observed so far, the scale at which new underlying physics could exist continues to rise, and the fine-tuning required to maintain the Higgs boson’s 125 GeV mass becomes increasingly pronounced.

This paper focuses on the hierarchy problem, which concerns explaining the low mass of the Higgs boson despite the large quantum corrections it should receive within the Standard Model. One possible solution is that the Higgs mass could vanish at tree level and only arise due to quantum corrections. Here, we explore the potential of flux compactification as a way forward.

Flux compactification plays a crucial role in both string theory and field theory \cite{Angelantonj:2002ct,Cremades:2004wa,Blumenhagen:2006ci}. It has been extensively studied due to its many attractive properties. For example, it can lead to a multiplicity of chiral fermions, provide an explanation for the number of lepton-quark generations \cite{Witten:1984dg}, break supersymmetry \cite{Bachas:1995ik}, and stabilize compact dimensions \cite{Braun:2006se}. In models with compact extra dimensions, it has been shown that zero modes of higher-dimensional gauge fields can act as the Higgs field in four dimensions \cite{Hosotani:1983xw,Hatanaka:1998yp,Dvali:2001qr}.\footnote{See \cite{Cremades:2004wa,Kobayashi:2010an,Hamada:2012wj} for other applications of magnetic compactification in phenomenology.} Identifying these zero modes with the Higgs scalar could result in a finite Higgs mass, protected by the size of the extra dimensions, with $m^2_{Higgs} \propto L^{-2}$ \cite{Arkani-Hamed:2001nha,Antoniadis:2001cv,Alfaro:2006is}. This mechanism becomes particularly compelling if the extra dimensions are large.

However, if the extra dimensions are small, a different solution is needed. Recent work has shown that, under magnetic compactification, quantum corrections to the mass of the zero mode of the scalar field, induced by additional components of the gauge field, can vanish. This was first demonstrated in a supersymmetric model in \cite{Buchmuller_2017}. A more detailed analysis of these vanishing one-loop corrections was conducted in \cite{Ghilencea:2017jmh}, with a focus on the regularization of divergent momentum integrals using dimensional regularization. In a follow-up study \cite{Buchmuller_2018}, the same authors explored the cancellation of one-loop corrections further, revealing that it stems from an exact shift symmetry in the higher-dimensional theory. Such shift symmetries were already known as a potential mechanism to protect the Higgs mass from UV corrections \cite{Peccei:1977hh,Panico:2015jxa}.

The one-loop corrections to the zero-mode mass have been calculated in a six-dimensional Yang-Mills theory with flux compactification \cite{Hirose:2019ywp}, confirming the cancellation of quantum corrections. Additionally, \cite{Hirose:2021rit} explored the possibility of generating a finite non-zero scalar mass and classified the interaction terms that break the shift symmetry. From a cosmological perspective, \cite{Hirose:2021xbs} suggested a new inflationary model in which the massless scalar is identified with the inflaton.

This paper is organized as follows. We begin with a brief review of flux compactification in Section \ref{sec:flux_comp_6D}, using a simple example of abelian gauge theory. We then introduce the non-supersymmetric model from \cite{Buchmuller_2018}, explaining how one-loop corrections to the scalar mass vanish due to a shift symmetry in the six-dimensional action.

In Section \ref{sec:two_U1}, we demonstrate how a $U(1) \times U(1)$ gauge symmetry results in a finite Higgs mass with vanishing one-loop corrections. This mechanism offers a way to protect the Higgs mass from large corrections even at high compactification scales.

Section \ref{sec:Tachyons} focuses on a six-dimensional Yang-Mills theory compactified on a torus with magnetic flux. In \cite{Buchmuller_2017}, tachyons were shown to appear in the $SU(2)$ model, indicating that the effective action represents an expansion around an extremal point that is not the true ground state. We propose a method to eliminate these tachyons by introducing a scalar boson charged under the gauge symmetry that acquires a vacuum expectation value.\footnote{As we will see, the elimination of tachyons is only valid for sufficiently low values of the magnetic flux.} Finally, we investigate whether this model could include chiral fermions.

\paragraph{Note added:} The first version of this paper was originally published on arXiv as a report on an internship at École Polytechnique, under the supervision of Emilian Dudas. This version is slightly shortened, with improved English and additional references. Since its publication, related work has appeared, including \cite{Akamatsu:2022jkg,Maru:2023esr,Kojima:2023umv,Hirose:2024vvx,Kojima:2024yog}. In particular, \cite{Kojima:2023umv} investigates the mass spectrum of a six-dimensional $SU(n)$ gauge theory with flux compactification, confirming the existence of tachyons in the model. This generalizes the discussion of Section \ref{sec:tac}. Moreover, \cite{Kojima:2024yog} presents an alternative approach to eliminating tachyonic modes, using flux compactification on a four-dimensional torus.

\section{Brief review of six-dimensional gauge theories with flux compactification}
\label{sec:flux_comp_6D}

In this section, we briefly review flux compactification on a torus for a $U(1)$ gauge theory and discuss the emergence of a Nambu-Goldstone boson resulting from symmetry-breaking in the extra dimensions. The results presented are primarily based on references \cite{Bachas:1995ik,Buchmuller_2018}, although similar frameworks have been widely explored in string theory flux compactifications, particularly in the context of supersymmetry breaking (see \textit{e.g.} \cite{Buchmuller:2016gib,Ghilencea:2017jmh,Angelantonj:2000hi}).

\subsection{Flux compactification on a torus}

Flux compactification creates a mass spectrum reminiscent of the Landau levels in quantum mechanics. Let us consider a scalar field with charge $q$ under an abelian gauge symmetry with coupling constant $g$, and associated with the gauge field $A^M$. The six-dimensional action for the scalar is
\begin{equation}
	S_6 = - \int d^6x D_M \overline{\chi} D^M \chi, 
\end{equation}
with
\begin{equation}
	D_M = \partial_M + igq A_M.
\end{equation}
We compactify the extra dimensions on a torus. For a factorizable torus, each of the compact dimensions is treated as a circle of length $2\pi r$. This corresponds to the periodic identification $y \leftrightarrow y+2\pi r$. In this case, the volume of the extra dimensions is $L^2=(2\pi r)^2$.
The magnetic flux background corresponds to a constant flux density $f$ in the internal dimensions. We make the following gauge choice:
\begin{equation}
	A_5 = -\frac{1}{2}fx_6 ,\qquad A_6=\frac{1}{2}fx_5 ,\qquad F_{56} = f.
	\label{eq:flux_background}
\end{equation}
For a torus of finite volume $L^2$, the flux is quantized,\footnote{This is shown in the appendix of \cite{Buchmuller_2018}. For another approach, see footnote 2 of \cite{Ghilencea:2017jmh}.}
\begin{equation}
	\frac{qg}{2\pi}\int_{T^2}F= \frac{qg}{2\pi}L^2f = N \in \mathbb{Z}.
\end{equation}
Without loss of generality, we choose $qf < 0$. By decomposing the kinetic term into a four-dimensional part and a part on $T^2$, one obtains
\begin{equation}
	\label{eq:flux_6D_action}
	S_6 = -\int \eta^{\mu\nu}D_{\mu}\overline{\chi}D_{\nu}\chi - \overline{\chi} H_2 \chi,
\end{equation}
where
\begin{equation}
	H_2 = -2qgf(a^{\dagger}a + 1/2)
\end{equation}
is written by defining creation and annihilation operators that depend on the internal space components:
\begin{align}
	\begin{split}
		&a = \frac{i}{\sqrt{-2qgf}}(\overline{\partial} - qgfz),\\
  &a^{\dagger} = \frac{i}{\sqrt{-2qgf}}(\partial +qgf\overline{z}),
	\end{split}
	\label{eq:ladder_operators}
\end{align}
with $z = \frac{1}{2}(x_5+ix_6)$ and $\partial = \frac{\partial}{\partial z} = \partial_5 -i\partial_6$. The ladder operators $a$, $a^{\dagger}$ satisfy the relation $[a,a^{\dagger}]=1$. We denote the internal fields as $\xi_{n,j}$, where $n$ is the Landeau level index and $j$ is the degeneracy index ranging from 0 to $|N|-1$. Starting from the lowest-mass fields,
\begin{equation}
	a\xi_{0,j} = 0 \qquad a^{\dagger}\overline{\xi}_{0,j}=0,
\end{equation}
one can define all the field profiles using the ladder operators:
\begin{equation}
	\xi_{n,j} = \frac{1}{\sqrt{n!}}\left(a^{\dagger} \right)^n\xi_{0,j} ,\qquad \overline{\xi}_{n,j} = \frac{1}{\sqrt{n!}}\left(a \right)^n\overline{\xi}_{0,j}.
	\label{eq:Landau_mode_functions}
\end{equation}
They satisfy the usual orthonormality relation.
\begin{equation}
	\int_{T^2}d^2x \xi_{n',j'}\overline{\xi}_{n,j} = \delta_{n,n'}\delta_{j,j'}.
	\label{eq:orthonormality_relation}
\end{equation}
The functions $\{\xi_{n,j}\}$ form a complete set of functions and charged fields can be expanded into to the Landau levels:
\begin{align}
\begin{split}
	&\chi(x_M) = \sum_{n,j} \chi_{n,j}(x_{\mu})\xi_{n,j}(x_m),\\
 &\overline{\chi}(x_M) = \sum_{n,j}\overline{\chi}_{n,j}(x_{\mu})\overline{\xi}_{n,j}(x_m).
	\label{eq:Landau_expansion}
\end{split}
\end{align}
Using the harmonic oscillator algebra, we can rewrite the six-dimensional action \eqref{eq:flux_6D_action} as
\begin{align}
\begin{split}
	S_4 = \int d^4x \sum_{n,j} &\left( -D_{\mu}\overline{\chi}_{n,j}D^{\mu}\chi_{n,j}+ 2qgf( n + 1/2 ) \overline{\chi}_{n,j}\chi_{n,j} \right),
\end{split}
\end{align}
which contains a mass term for the scalars $\chi_{n,j}$ \cite{Bachas:1995ik}:
\begin{equation}
	m_{n,j} = -2qgf(n+1/2) = \frac{2\pi|N|}{L^2}(n+1/2).
\end{equation}

\subsection{Fermion with abelian flux background}
\label{sec:nonsusy_model}

In this section, we introduce here a six-dimensional fermion charged under an abelian gauge theory with a magnetic flux present in the background. This model was presented in detail in \cite{Buchmuller_2018}. The six-dimensional Lagrangian for a single fermion with abelian gauge symmetry is
\begin{equation}
	S_6 = \int d^6x \left( -\frac{1}{4}F^{MN}F_{MN} + i \overline{\Psi}\Gamma^MD_M\Psi \right),
	\label{eq:6D_nonsusy_action}
\end{equation}
where $D_M = \partial_M +iqA_M$. The six-dimensional Weyl spinor is split into two two-component Weyl spinors $\psi$ and $\chi$ with charges $q$ and $-q$, respectively.
\begin{align}
	\begin{split}
		\Psi = \begin{pmatrix} \psi_L\\ \psi_R \end{pmatrix}, \quad \psi_L &= \begin{pmatrix} \psi \\ 0 \end{pmatrix}, \quad \psi_R = \begin{pmatrix} 0 \\ \overline{\chi} \end{pmatrix},\\
		\gamma_5\psi_L = - \psi_L ,&\quad \gamma_5\psi_R = \psi_R.
		\label{eq:convention_6D_fermion}
	\end{split}
\end{align}
We choose the following basis for the gamma matrices $\Gamma^M$:
\begin{align}
\begin{split}
	&\Gamma^{\mu} = \begin{pmatrix} \gamma^{\mu} & 0 \\ 0 & \gamma^{\mu} \end{pmatrix} ,\\
 &\Gamma^5 = \begin{pmatrix} 0 & i\gamma_5 \\ i\gamma_5 & 0 \end{pmatrix} ,\quad 
 \Gamma^{6} = \begin{pmatrix} 0 & -\gamma_5 \\ \gamma_5 & 0 \end{pmatrix},
	\label{eq:6D_gamma}
\end{split}
\end{align}
which satisfy the algebra $\{\Gamma_M,\Gamma_N\} = -2 \eta_{MN}$. Expanding the fermionic part of the action (\ref{eq:6D_nonsusy_action}), one finds
\begin{align}
\begin{split}
	S_{6f}= \int d^6x &\left( -i\psi \sigma^{\mu}\overline{D}_{\mu}\overline{\psi} - i\chi\sigma^{\mu}D_{\mu}\overline{\chi} - \chi( \partial + \sqrt{2}q\phi )\psi - \overline{\chi}(\overline{\partial}+\sqrt{2}q\overline{\phi})\overline{\psi}  \right),\label{eq:non_susy_fermionic_action}
\end{split}
\end{align}
where $D_{\mu}=\partial_{\mu}+iqA_{\mu}$, $\overline{D}_{\mu}=\partial_{\mu}-iqA_{\mu}$, and
\begin{equation}
	\phi=\frac{1}{\sqrt{2}}(A_6+iA_5).
\end{equation}
Similarly, the gauge part of the action is
\begin{align}
\begin{split}
	S_{6g} &= \int d^6x \left( -\frac{1}{4}F^{\mu\nu}F_{\mu\nu} - \partial^{\mu}\overline{\phi}\partial_{\mu}\phi - \frac{1}{4}(\partial\overline{\phi} + \overline{\partial}\phi)^2 \right.\\
 &\left.- \frac{1}{2} \overline{\partial}A^{\mu}\partial A_{\mu} - \frac{i}{\sqrt{2}}\partial_{\mu}A^{\mu}(\partial\overline{\phi} - \overline{\partial\phi})\right).
	\label{eq:non_susy_bosonic_action}
\end{split}
\end{align}
We compactify on a torus $T^2$ with a quantized magnetic flux in the background (\ref{eq:flux_background}). We assume without loss of generality that $qf>0$. The ladder operators take the form
\begin{align}
	\begin{split}
		a_+^{\dagger} &= \frac{i}{\sqrt{2qf}}(\overline{\partial} - qfz) ,\qquad a_+ = \frac{i}{\sqrt{2qf}}(\partial +qf\overline{z}),\\
		a_-^{\dagger} &= \frac{i}{\sqrt{2qf}}(\partial - qf\overline{z}), \qquad a_- = \frac{i}{\sqrt{2qf}}(\overline{\partial} + qfz),
		\label{eq:ladder_operators_non_susy}
	\end{split}
\end{align}
and satisfy the relations $[a_{\pm},a^{\dagger}_{\pm}]=1$ and $ [a_{\pm},a_{\pm}] = [a_{\pm},a^{\dagger}_{\mp}]=0$. We define the orthonormal set of wave functions as
\begin{align}
	\begin{split}
		a_+\xi_{0,j} = 0, &\qquad a_-\overline{\xi}_{0,j}=0 ,\\
		\xi_{n,j} = \frac{i^n}{\sqrt{n\!}}(a_+^{\dagger})^n\xi_{0,j} ,&\qquad \overline{\xi}_{n,j} = \frac{i^n}{\sqrt{n\!}}(a_-^{\dagger})^n\overline{\xi}_{0,j},
		\label{eq:mode_functions_definition}
	\end{split}
\end{align}
where $j$ takes values between $0$ and $|N|-1$. The Weyl spinors are expanded with respect to these mode functions as
\begin{equation}
		\psi = \sum_{n,j} \psi_{n,j} \xi_{n,j},\quad \chi = \sum_{n,j} \chi_{n,j} \overline{\xi}_{n,j}.
\end{equation}
In contrast, the gauge fields do not feel the magnetic flux and thus expand with respect to KK mode functions. Note that the fermionic zero modes are chiral. Indeed, starting from the Lagrangian (\ref{eq:non_susy_fermionic_action}), one can write down the Euler-Lagrange equations for the Weyl fermions. We now compute the equation of motion for $\psi$ as an example:
\begin{align}
	\begin{split}
		\mathscr{L}_f &\supset - i\psi\sigma^{\mu}\partial_{\mu}\overline{\psi} - \chi(\partial+qf\overline{z})\psi,\\
  &\Rightarrow 
		- i \sigma^{\mu}\partial_{\mu}\overline{\psi} - i\sqrt{2qf}a_-^{\dagger}\chi = 0,
	\end{split}
\end{align}
where we have expanded the scalar field $\phi$ around its vacuum configuration due to the constant magnetic field:
\begin{equation}
	\phi = \frac{f}{\sqrt{2}}\overline{z} + \varphi,
\end{equation}
and performed partial integration to isolate $\psi$. By performing a similar calculation for $\chi,\overline{\chi},\overline{\psi}$ and combining the coupled equations, we find
\begin{align}
	\begin{split}
		\Box \psi - \mathcal{M}_+^2 \psi &=0,\\
		\Box \chi - \mathcal{M}_-^2 \chi &=0,\\
	\end{split}
\end{align}
where
\begin{align}
	\begin{split}
		\mathcal{M}_+^2 &= 2qfa_+^{\dagger}a_+, \\
		\mathcal{M}_-^2 &= 2qf(a_-^{\dagger}a_-+1),
	\end{split}
	\label{eq:mass_operators}
\end{align}
and we have used the property $\sigma^{\mu}\overline{\sigma}^{\nu}\partial_{\mu}\partial_{\nu} = \Box$. Hence, there are $|N|$ left-handed fermionic zero modes.

Using the decomposition in Landau and KK modes, the four-dimensional effective Lagrangian of the theory reads\footnote{Again, we only include the zero modes of the uncharged fields $A_{\mu},\varphi$.}
\begin{align}
	\begin{split}
		S_4 &= \int d^4x \Bigg[ -\partial^{\mu}\overline{\varphi}_0\partial_{\mu}\varphi_0 + \sum_{n,j} \Bigg(  i \overline{\psi}_{Lj}\gamma^{\mu}D_{\mu}\psi_{Lj} +  \Bigg. \Bigg. \\
		&i\overline{\Psi}_{n,j}\gamma^{\mu}D_{\mu}\Psi_{n,j} + \sqrt{2qf(n+1)}\overline{\Psi}_{n,j}\Psi_{n,j} \\
  &+ \sqrt{2}q\varphi_0 \left( \overline{\Psi}_{0,j}\frac{1-\gamma_5}{2}\psi_{Lj} + \overline{\Psi}_{n+1,j}\frac{ \gamma_5-1}{2}\Psi_{n,j} \right) \\
		& + \Bigg. \Big. \sqrt{2}q\overline{\varphi_0} \left( \overline{\psi}_{Lj} \frac{1+\gamma_5}{2}\Psi_{0,j} + \overline{\Psi}_{n,j}\frac{1+\gamma_5}{2}\Psi_{n+1,j} \right) \Bigg) \Bigg],
		\label{eq:nonsusy_action_flux}
	\end{split}
\end{align}
where $\psi_{Lj}$ are $|N|$ left-handed fermions and $\Psi_{n,j}$ is an infinite tower of massive Dirac fermions
\begin{equation}
	\psi_{Lj}=\begin{pmatrix} \psi_{0,j} \\ 0 \end{pmatrix} ,\qquad \Psi_{n,j}=\begin{pmatrix} \psi_{n+1,j} \\ \overline{\chi}_{n,j} \end{pmatrix}.
\end{equation}
The zero mode of the scalar, often called the Wilson line (WL), is massless. We explain below that the vanishing of the WL scalar mass is due to a continuous shift symmetry. 

\subsection{Quantum corrections}

\begin{figure}[h]
	\centering
	\includegraphics[width=8cm]{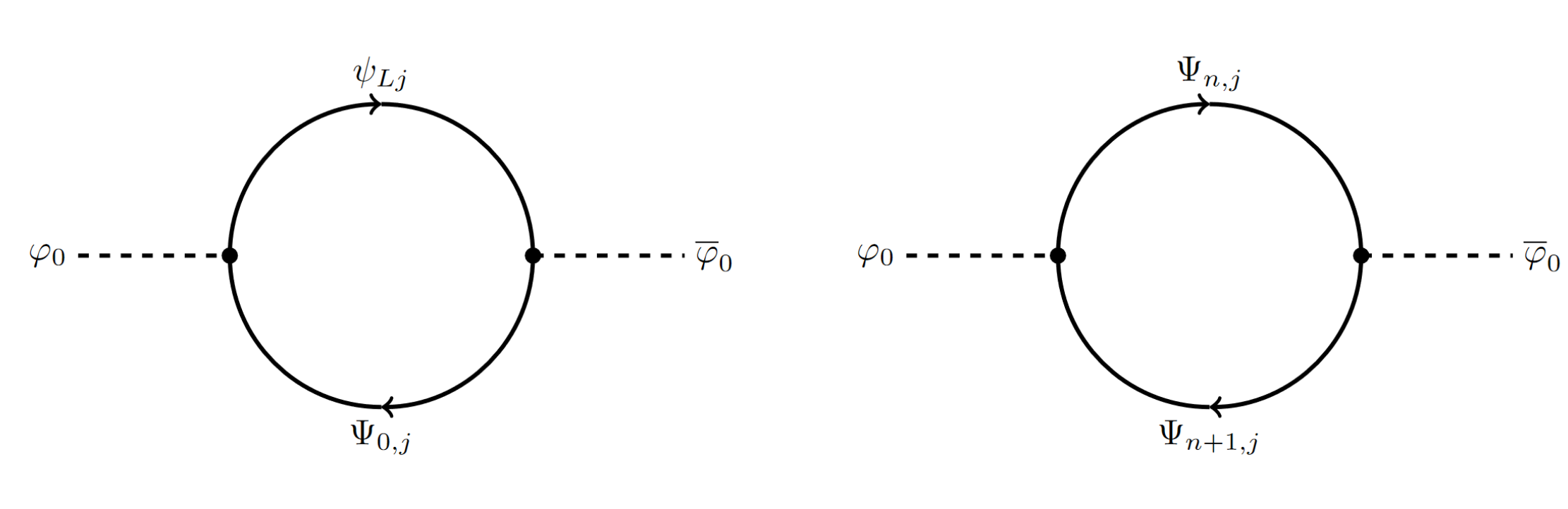}
	\caption{One-loop contributions to the WL mass.}
	\label{fig:nonsusy_mass_correction_flux}
\end{figure}

The action (\ref{eq:nonsusy_action_flux}) leads to two fermionic contributions to the corrections to the WL scalar mass, as shown in Figure \ref{fig:nonsusy_mass_correction_flux}. 

From these diagrams, we obtain the following corrections:
\begin{align}
	\begin{split}
		\delta m_{1}^2 &= -2q^2|N|\sum_{n}\int \frac{d^4k}{(2\pi)^4} \frac{2k^2}{(k^2+2qfn)(k^2+2qf(n+1))}\\
		&= 4q^2|N|\sum_{n}\int \frac{d^4k}{(2\pi)^4} \left( \frac{n}{k^2+2qfn} - \frac{n+1}{k^2+2qf(n+1)} \right)\\
		&= \frac{q^2}{4\pi^2}|N| \int_0^{\infty} dt \frac{1}{t^2} \sum_n \left(ne^{-2qnft} - (n+1)e^{-2qf(n+1)}\right)\\
		&= \frac{q^2}{4\pi^2}|N| \int_0^{\infty} dt \frac{1}{t^2} \left( \frac{e^{2qft}}{(e^{2qft}-1)^2}-\frac{e^{2qft}}{(e^{2qft}-1)^2}\right)\\
		&=0,
		\label{eq:mass_correction_computation_nonsusy}
	\end{split}
\end{align}
where we have used the Schwinger representation of the propagators. We investigate the origin of this seemingly miraculous cancellation in the next section. The cancellation of one-loop corrections in a Yang-Mills $SU(2)$ non-supersymmetric model was previously verified in \cite{Hirose:2019ywp}.

\subsection{Underlying symmetry}

\label{sec:flux_symmetry}

The actions (\ref{eq:non_susy_fermionic_action}) and (\ref{eq:non_susy_bosonic_action}) are invariant under translation $\delta_T = \epsilon\partial + \overline{\epsilon}\overline{\partial}$ on the torus $T^2$, acting on $\chi,\psi,A^{\mu}$. However, the magnetic flux in the background explicitly breaks the translational symmetry, which can be compensated by a shift in the scalar fluctuation $\varphi$,
\begin{equation}
	\delta_T \varphi = (\epsilon\partial + \overline{\epsilon}\overline{\partial})\varphi + \frac{\overline{\epsilon}}{\sqrt{2}}f.
\end{equation}
The translation generators do not commute with the mass operators (\ref{eq:mass_operators}) so that the effect of translation on the mode functions cannot be trivial. Instead, the whole tower of fields is reshuffled.

To construct a general transformation of the mode functions, one combines the translation $\delta_T = \epsilon\partial + \overline{\epsilon}\overline{\partial}$ with another symmetry of the action:
\begin{align}
\begin{split}
	&\varphi_{\Lambda} = -\frac{1}{\sqrt{2}}\partial \Lambda ,\quad \psi_{\Lambda} =e^{q\Lambda}\psi,\\
 &\chi_{\Lambda} = e^{-q\Lambda}\chi, \quad \Lambda = f(\alpha \overline{z} - \overline{\alpha}z),
 \end{split}
\end{align}
where $\alpha$ is a complex parameter. These transformations were first considered in \cite{Scherk:1978ta}. For the WL, the transformation corresponds to a shift $\delta_{\Lambda} \varphi = \frac{\overline{\alpha}}{\sqrt{2}}f$. Combining the two symmetries of the Lagrangian, one obtains the infinitesimal transformation
\begin{equation}
	\delta \psi = (\delta_T + \delta_{\Lambda}) \psi = -i\sqrt{2qf}(\epsilon a_+ +  \epsilon a_+^{\dagger})\psi.
\end{equation}
This transformation reshuffles the different modes, connecting neighboring mode functions:
\begin{align}
	\begin{split}
		\delta \psi &= \sum_{n,j} \delta \psi_{n,j} \xi_{n,j}, \\
		\delta \psi_{n,j} &:= \sqrt{2qf} (\epsilon\sqrt{n+1}\psi_{n+1,j} - \overline{\epsilon}\sqrt{n}\psi_{n-1,j}).
	\end{split}
\end{align}
Analogously,
\begin{align}
	\begin{split}
		\delta \chi &= \sum_{n,j} \delta \chi_{n,j} \xi_{n,j}, \\
		\delta \chi_{n,j} &:= \sqrt{2qf} (\overline{\epsilon}\sqrt{n+1}\chi_{n+1,j} - \epsilon\sqrt{n}\chi_{n-1,j}).
	\end{split}
\end{align}
The invariance of (\ref{eq:nonsusy_action_flux}) under these transformations is explicitly verified in \cite{Buchmuller_2018}, under the condition that $\varphi$ transforms as
\begin{equation} \delta \varphi_0 = \sqrt{2}\overline{\epsilon}f. \label{eq
} \end{equation}
In the context of a $U(1)$ gauge theory, as discussed here, the effective Lagrangian also possesses an exact symmetry under which the scalar zero modes transform according to the shift in (\ref{eq
}). This allows the WL scalar to be identified as the Nambu-Goldstone boson corresponding to the translational symmetry on the torus \cite{Buchmuller_2017,Honda:2019ema}.

\section{A scalar of finite and small mass in $U(1)\times U(1)$ gauge theory}

\label{sec:two_U1}

One of the motivations for the construction presented above is to construct toy models of the Higgs boson with finite mass. To do so, it is necessary to break the translational symmetry to generate a finite mass of the TeV order for the Higgs boson. Rather than adding ad-hoc interaction terms as in\cite{Hirose:2021rit}, we would like to investigate the possibility of generating non-vanishing finite corrections to the scalar mass using only the gauge structure of the theory. It has been mentioned in \cite{Buchmuller_2017} that in the case of a gauge group with several $U(1)$ gauge factors, the WLs are the pseudo-Nambu-Goldstone bosons of the translational symmetry. We develop this idea further in this section.

\subsection{Pseudo-Nambu-Goldstone bosons}

In the $U(1$) case, the WL can be seen as the Nambu-Goldstone boson of the translational symmetry on the torus. If the gauge symmetry contains more than one $U(1)$, the situation becomes more subtle. In contrast to the discussion in \cite{Buchmuller_2017}, we discuss this idea in the non-supersymmetric model of \cite{Buchmuller_2018} presented in Section \ref{sec:nonsusy_model}. We study the simplest case of a $U(1)_1\times U(1)_2$ gauge symmetry. There are two gauge fields and therefore two WL scalars denoted $\varphi^1$ and $\varphi^2$. The covariant derivative reads
\begin{equation}
	D_M  = \partial_M + iq_{\alpha}A_M^{\alpha} ,
\end{equation}
where $q_{\alpha}$ and $A^{\alpha}_M$ are the charge of $\Psi$ and the gauge vector field associated with $U(1)_{\alpha}$. Additionally,
\begin{equation}
	A_5^{\alpha} = -\frac{1}{2}f^{\alpha}x_6, \quad A_6^{\alpha} = \frac{1}{2}f^{\alpha}x_5, \qquad F_{56}^{\alpha} = f^{\alpha}.
\end{equation}
If one includes an arbitrary number of fermion families $N_f$, the charge would take the form of a matrix $q_{i\alpha}$.

Following the same procedure as in Section \ref{sec:flux_symmetry}, if the Lagrangian transforms into a total derivative, the WLs transform as
\begin{equation}
	q_{\alpha} \delta \varphi_0^{\alpha} = \sqrt{2}q_{\alpha}\overline{\epsilon}f^{\alpha}.
\end{equation}
This indicates that a field charged under two $U(1)$ gauge symmetries should feel an effective flux depending on the two charges and the flux of the gauge groups $U(1)_{\alpha}$. For example, consider two fermions doublets $\{\Psi^i\}_{i=1,2}$ with charges
\begin{equation}
	q_{i\alpha} = \begin{pmatrix} 1 & 1 \\ -1 & 1 \end{pmatrix}.
\end{equation}
For the flux choice $f^{1}=f^2=f$, one has
\begin{equation}
	q_{i\alpha}\delta \varphi_0^{\alpha} = \sqrt{2} f \begin{pmatrix} 2\overline{\epsilon}^1 \\ 0 \end{pmatrix}.
\end{equation}
In this case, the second fermion $\Psi^2$ does not feel any flux. Additionally, there is one Nambu-Goldstone boson corresponding to $\varphi_0^1+\varphi_0^2$ while $\varphi_0^1-\varphi_0^2$ does not transform non-linearly. Hence, there is only one Nambu-Goldstone boson. 

This is a general feature of the theory. There are two translational symmetries in the extra space: one along $x_5$ and the other along $x_6$. When at least one of the gauge fields has flux in its background, both of these symmetries are broken and a complex Nambu-Goldstone boson arises. Since no other symmetries are broken beyond these translational symmetries, any additional symmetries would be accidental and potentially broken by interaction terms. Therefore, one can have at most one Nambu-Goldstone boson, and any other Nambu-Goldstone boson would be a "pseudo-Nambu-Goldstone boson" with a non-vanishing mass. 

\subsection{Vanishing of scalar masses to leading order}

\begin{figure}[ht]
	\centering
	\includegraphics[width=8cm]{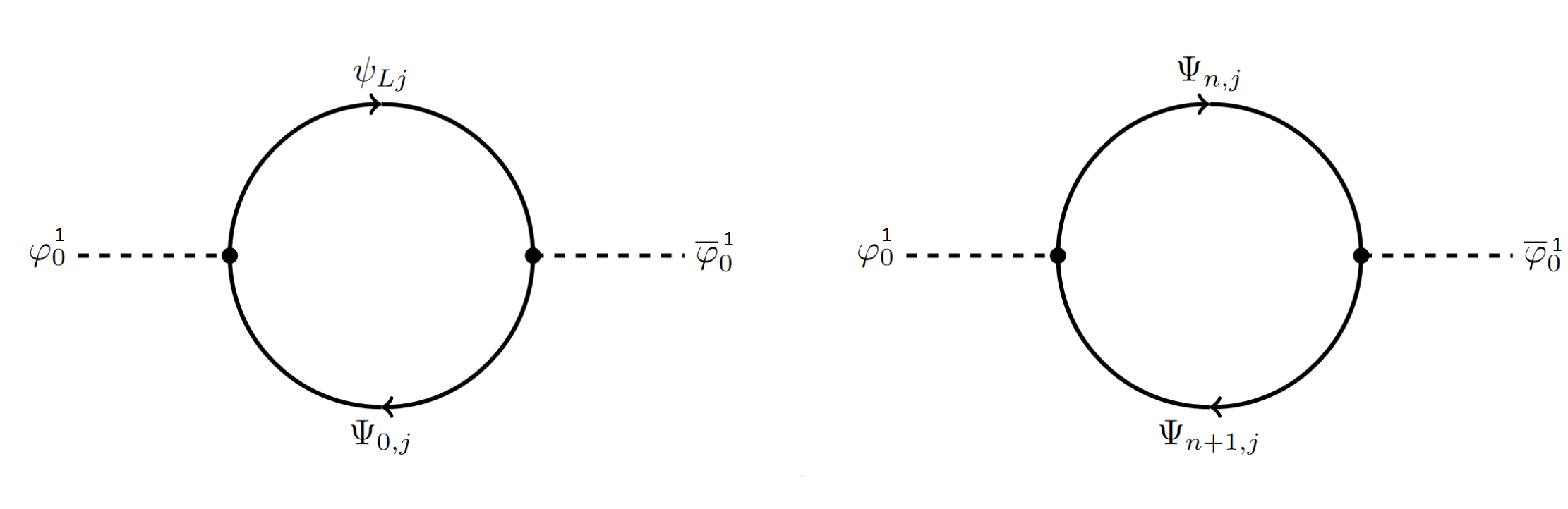}
	\includegraphics[width=8cm]{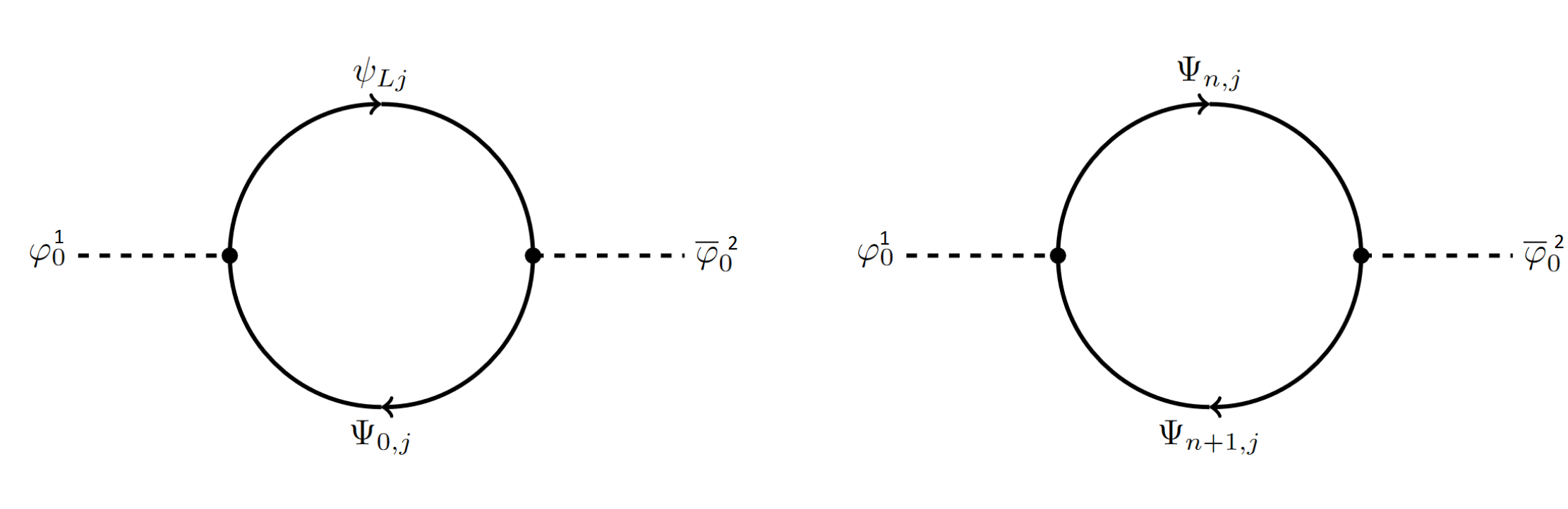}
	\includegraphics[width=8cm]{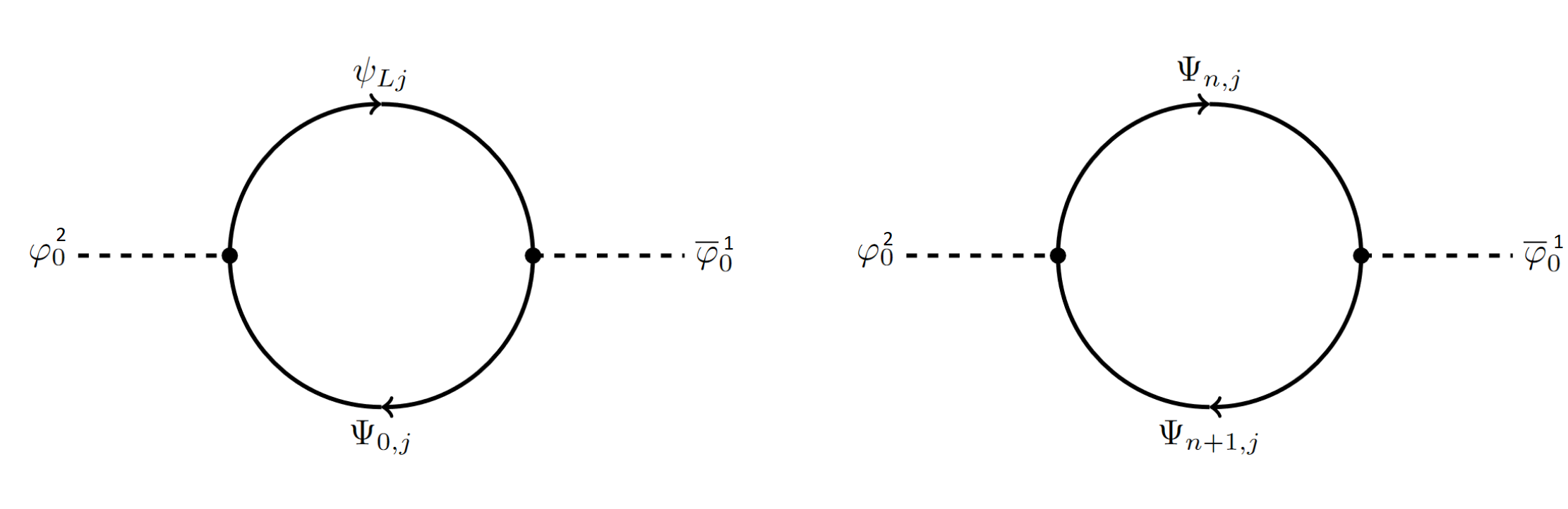}
	\includegraphics[width=8cm]{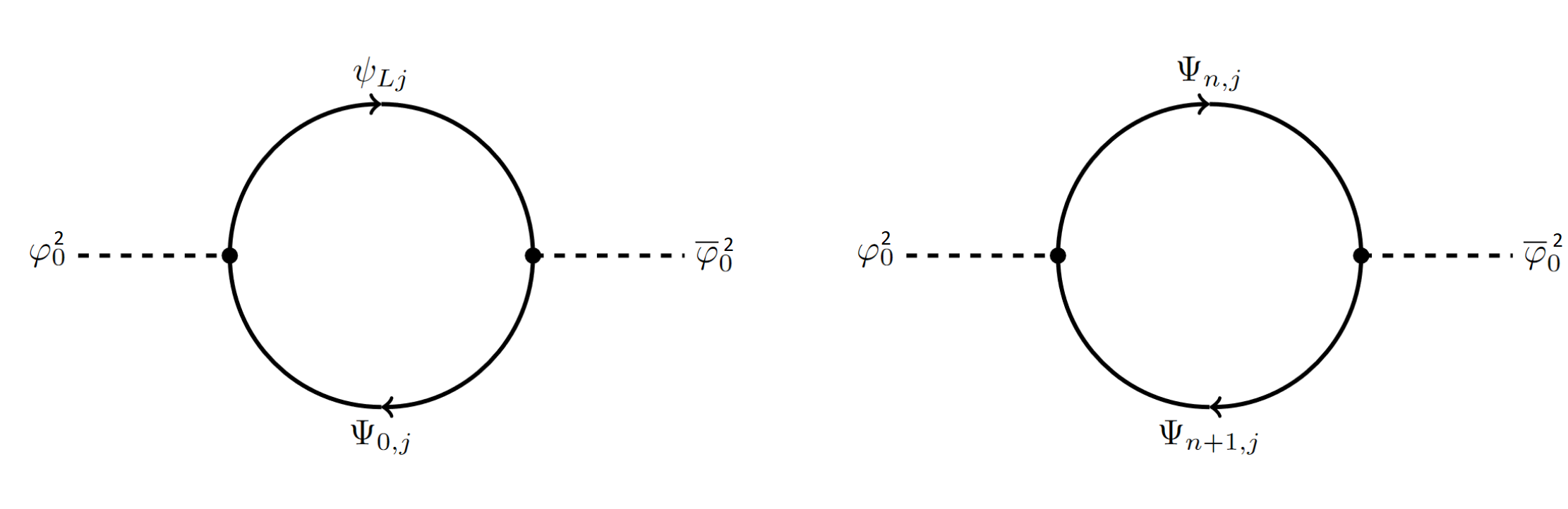}
	\caption{One-loop contributions to the WL masses for $N_f=1$.}
	\label{fig:nonsusy_mass_correction_flux_twoU1}
\end{figure}

Although the mass of a pseudo-Nambu-Goldstone boson is finite, it should be small compared to a standard boson mass. We repeat the calculations of Section \ref{sec:nonsusy_model} with two $U(1)$ groups instead of one and verify that the effective action has the same form as (\ref{eq:nonsusy_action_flux}) with $q\varphi_0$ replaced by $q_{\alpha}\varphi_0^{\alpha}$. The computation easily generalizes to an arbitrary number $N_f$ of fermion families.
\begin{align}
	\begin{split}
		\mathscr{L}_4 &\supset \sum_{i=1}^{N_f} \sum_{n\geq 0} \sum_{j=0}^{N_i-1} \Bigg( \sqrt{2}q_{i\alpha}\varphi_0^{\alpha} \left( \overline{\Psi}_{0,j}^i\frac{1-\gamma_5}{2}\psi_{Lj}^i + \overline{\Psi}^i_{n+1,j}\frac{ \gamma_5-1}{2}\Psi^i_{n,j} \right)  \Bigg. \\
        &\Bigg. + \sqrt{2}q_{i\alpha}\overline{\varphi_0}^{\alpha} \left( \overline{\psi}^i_{Lj} \frac{1+\gamma_5}{2}\Psi^i_{0,j}   + \overline{\Psi}^i_{n,j}\frac{1+\gamma_5}{2}\Psi^i_{n+1,j} \right) \Bigg),
		\label{eq:lagrangian_two_U1}
	\end{split}
\end{align}
where $N_i$ is the quantization of the flux associated with the charge $i$,
\begin{equation}
	\frac{q_{i\alpha}}{2\pi}f^{\alpha} = N_i \in \mathbb{Z}.
\end{equation}
With the interaction Lagrangian above, we compute the one-loop correction to the WL masses. Here, because of the multitude of WLs, one has to consider possible mixing between the two scalars. Indeed, there are eight one-loop diagrams to consider, see Figure \ref{fig:nonsusy_mass_correction_flux_twoU1}.
For an arbitrary number of families $N_f$, we find
\begin{align}
	\begin{split}
		(\delta m_1^2)_{\alpha\beta} &= -2 \sum_{i=1}^{N_f} q_{i\alpha}q_{i\beta} |N_i| \sum_{n} \int \frac{d^4k}{(2\pi)^4} \frac{2k^2}{(k^2+2q_{i\gamma}f^{\gamma}n)(k^2+2q_{i\nu}f^{\nu}(n+1))} \\
		&= 4\sum_{i=1}^{N_f} q_{i\alpha}q_{i\beta}|N_i|\sum_n \int \frac{d^4k}{(2\pi)^4}\left( \frac{n}{k^2+2q_{i\gamma}f^{\gamma}n} - \frac{n+1}{k^2+2q_{i\nu}f^{\nu}(n+1)} \right)\\
		&=0,
	\end{split}
\end{align}
if $q_{i\alpha}f^{\alpha}\neq 0$. Thus, one-loop corrections to WLs scalar masses always vanish when the effective flux is non-zero.

This confirms that the pseudo-NG boson masses are protected from one-loop corrections and one could hope to produce a realistic mass in the TeV range. Further calculations at two-loop order would be needed to compute the precise spectrum of the WLs.

\subsection{Two-loop corrections from interaction terms}

If finite corrections are expected, one should explicitly observe symmetry-breaking terms in the Lagrangian. The transformations of fermions Landau modes found in \cite{Buchmuller_2017} can be generalized to the $U(1)_1\times U(1)_2$ case with $N_f$ families:
\begin{align}
	\begin{split}
		\delta \psi_{n,j}^i&= \sqrt{2q_{i\alpha}f^{\alpha}}\left( \epsilon^i\sqrt{n+1}\psi^i_{n+1,j} - \overline{\epsilon}^i\sqrt{n}\psi^i_{n-1,j} \right), \\
		\delta \chi_{n,j}^i&= \sqrt{2q_{i\alpha}f^{\alpha}}\left( -\epsilon^i\sqrt{n}\chi^i_{n-1,j} + \overline{\epsilon}^i\sqrt{n+1}\chi^i_{n+1,j} \right).
		\label{eq:transf_law_fermions_two_U1}
	\end{split}
\end{align}
The variation of the mass term is
\begin{align}
	\begin{split}
		\delta &\left( - \sum_{n,j}\sqrt{2q_{i\alpha}f^{\alpha}(n+1)} \chi^i_{n,j}\psi^i_{n+1,j}\right) \\
	 &=   2\overline{\epsilon}^i q_{i\alpha}f^{\alpha} \sum_{n,j} \chi^i_{n,j}\psi^i_{n,j},
	\end{split}
\end{align}
which must compensate for the variation of the Yukawa term
\begin{align}
	\begin{split}
		&\delta \left( - \sum_{n,j} \sqrt{2}q_{i\alpha}\varphi_0^{\alpha}\chi_{n,j}^i\psi^i_{n,j} \right)\\
		&= -\sqrt{2}q_{i\alpha}\delta (\varphi_0^{\alpha})\sum_{n,j}\chi_{n,j}^i\psi^i_{n,j}.
	\end{split}
\end{align}
These equations impose the following transformation law for the WLs,
\begin{equation}
	q_{i\alpha}\delta\varphi_0^{\alpha} = \sqrt{2} \overline{\epsilon}^{i} q_{i\alpha}f^{\alpha}, 
	\label{eq:trans_law_WL_two_U1}
\end{equation}
matching the transformation law mentioned at the beginning of this section. This transformation cannot leave the full Lagrangian invariant. However, the $N_f$ shift symmetries are preserved by the Yukawa interaction, which explains why the one-loop correction vanishes. We now show that the interaction term involved in two-loop corrections breaks the $N_f$ shift symmetries only to preserve one particular shift transformation. The transformation law of massive modes of gauge fields reads
\begin{align}
	\begin{split}
		\delta \varphi_{l,m}^{\alpha} &= \left( \epsilon^{\alpha} M_{l,m} - \overline{\epsilon}^{\alpha}\overline{M}_{l,m} \right) \varphi_{l,m}^{\alpha}, \\
		\delta A_{\mu;l,m}^{\alpha} &= \left( \epsilon^{\alpha} M_{l,m} - \overline{\epsilon}^{\alpha}\overline{M}_{l,m} \right) A_{\mu;l,m}^{\alpha},
		\label{eq:transf_law_massive_gauge_two_U1}
	\end{split}
\end{align}
where we introduced two different parameters $\epsilon^{\alpha}$ for each set of gauge fields. Let us consider as an example an interaction term between massive gauge fields found in \cite{Buchmuller_2018} and extend it to the $U(1)\times U(1)$ case:
\begin{equation}
	\mathscr{L}_4 \supset -\sqrt{2} \sum_{i=1}^{N_f}\sum_{l,m;n,j;n',j'} C_{n,j;n',j'}^{l,m} q_{i\alpha}\varphi_{l,m}^{\alpha}\chi_{n,j}^i\psi_{n,j}^i,
\end{equation}
where
\begin{equation}
	C_{n,j;n',j'}^{l,m} = \int_{T^2}d^2x e^{zM_{l,m}-\overline{z}\overline{M}_{l,m}}\overline{\xi}_{n,j}\xi_{n',j'}.
\end{equation}
The transformation of this term under (\ref{eq:transf_law_massive_gauge_two_U1}) reads
\begin{align}
	\begin{split}
    \sum_{i=1}^{N_f}\sum_{l,m;n,j;n',j'} &\sqrt{2} q_{i\alpha} \big((\overline{\epsilon}^{\alpha}-\overline{\epsilon}^i)\overline{M}_{l,m} - (\epsilon^{\alpha}-\epsilon^i)M_{l,m}\big) C_{n,j;n',j'}^{l,m} \varphi_{l,m}^{\alpha}\chi_{n,j}^i\psi_{n,j}^i,
	\end{split}
\end{align}
which is non-vanishing in general. The only case leading to a trivial transformation is
\begin{equation}
	\epsilon^i = \epsilon^{\alpha} \equiv \epsilon \qquad \forall i ,\alpha.
\end{equation}
Hence, there is only one parameter for the shift symmetry of the WLs — only one shift symmetry. The only linear combination of the WLs transforming with a shift and hence being the NG boson is
\begin{equation}
	\frac{\varphi^1_0 + \varphi_0^2}{\sqrt{2}}.
\end{equation}

\section{Six-dimensional Yang-Mills theories with flux compactification}

\label{sec:Tachyons}

To develop a phenomenologically relevant effective theory, it is essential to extend the model to Yang-Mills gauge theories. It was shown in \cite{Buchmuller_2017} that the effective action for non-abelian flux contains tachyons. Indeed, the mass shift for component of a six-dimensional multiplet with flux compactification is \cite{Bachas:1995ik}
\begin{equation}
	\delta \mathcal{M}_q^2 = (2n+1)|qgf| + 2qgf \Sigma, 
	\label{Bachas_equation}
\end{equation}
where $\Sigma$ is the internal helicity, equal to $\pm 1/2$ for fermions, $0$ for scalars and $A^{\mu}$, and $\pm 1$ for $A_{5,6}$. Note that this mass equation does not predict the entire spectrum found in \cite{Buchmuller_2017} as some of the fields are absorbed via the Stückelberg mechanism.

The abelian flux background of Section \ref{sec:nonsusy_model} is perturbatively stable. For a non-abelian flux background, the effective action is derived from an expansion around a local maximal point rather than a ground state, and some of the extra-dimensional gauge fields become tachyonic. The usual solution is to study tachyon condensation to reveal the true ground state of the theory. See \cite{Hashimoto:2003xz,Sen:2004nf} for applications in string theory. In this section, we note that the issue can be resolved by the presence of additional fields in the scalar mass spectrum, while maintaining the chirality of fermions.

\subsection{Tachyons in Yang-Mills theories with flux compactification}

\label{sec:tac}
In this section, we compute the scalar spectrum of a model with a $SU(2)$ gauge symmetry with flux compactification on a torus and show that it contains tachyonic modes. We consider the kinetic Lagrangian for a six-dimensional vector field
\begin{equation}
	\mathscr{L}_6 = - \frac{1}{2} \Tr{F_{MN}F^{MN}},
	\label{eq:kinetic_gauge_Lag}
\end{equation}
where the fields are in the adjoint representation of $SU(2)$ and develop on the basis of the $SU(2)$ algebra, which consists of the Pauli matrices $\{T_1,T_2,T_3\}$ normalized such that $\Tr T_aT_b = \frac{1}{2}\delta_{ab}$. For non-abelian gauge theories,
\begin{equation}
	F_{MN}=\partial_MA_N-\partial_NA_M-ig[A_M,A_N].
\end{equation}
To find the four-dimensional effective action, we split the six-dimensional space into four dimensions and two additional dimensions compactified on a torus $T^2$. The associated electromagnetic tensor is written as
\begin{align}
	\begin{split}
		F_{56} &= \frac{1}{\sqrt{2}}(\partial\overline{\phi}+\overline{\partial}\phi-\sqrt{2}g[\phi,\overline{\phi}]),\\
		F_{\mu5}^2+F_{\mu6}^2 &= (F_{\mu6}+iF_{\mu5})(F^{\mu6}-iF^{\mu5}),\\
		F_{\mu6}+iF_{\mu5} &= \sqrt{2}\partial_{\mu}\phi - i\partial A_{\mu} -i\sqrt{2}g[A_{\mu},\phi].
	\end{split}
\end{align}
Equation (\ref{eq:kinetic_gauge_Lag}) can be rewritten as
\begin{align}
	\begin{split}
		\mathscr{L}_6 &= -\frac{1}{2}\Tr{F_{\mu\nu}F^{\mu\nu}}\\
        &-2\Tr\left( \partial_{\mu}\overline{\phi} - \frac{i}{\sqrt{2}}\overline{\partial} A_{\mu} -i\sqrt{2}g[A_{\mu},\overline{\phi}] \right)\left( \partial_{\mu}\phi - \frac{i}{\sqrt{2}}\partial A_{\mu} -ig[A_{\mu},\phi] \right)\\
		&- \frac{1}{2}\Tr\left(\partial\overline{\phi}+\overline{\partial}\phi-\sqrt{2}g[\phi,\overline{\phi}]\right)\left(\partial\overline{\phi}+\overline{\partial}\phi-\sqrt{2}g[\phi,\overline{\phi}]\right).
	\end{split}
\end{align}
The mass terms of the scalar are contained in the third line of the Lagrangian. 

To formulate the theory in terms of charged components of the gauge fields, one defines a new basis of generators ${T_+,T_-,T_3}$ where
\begin{equation}
	T_{\pm} = T_1 \pm i T_2.
	\label{eq:def_basis_algebra}
\end{equation}
The basis elements satisfy
\begin{align}
	\begin{split}
		&\Tr{T_3^2}=\frac{1}{2}, \quad \Tr{T_+T_-}=1, \quad \Tr{T_{\pm}T_3}=0,\\
 & \Tr{T_{\pm}^2}=0,\quad 
		[T_+,T_-]=2T_3, \quad [T_3,T_{\pm}]=\pm T_{\pm}.
	\end{split}
	\label{eq:traces_commutations}
\end{align}
An arbitrary field $\phi$ in the adjoint representation is decomposed as
\begin{align}
	\phi &= \phi_3 T_3 + \phi_+ \frac{T_-}{\sqrt{2}} + \phi_- \frac{T_+}{\sqrt{2}},\\
	\overline{\phi} &= \overline{\phi}_3 T_3 + \overline{\phi}_+\frac{T_+}{\sqrt{2}} + \overline{\phi}_-\frac{T_-}{\sqrt{2}}.
\end{align}
Vector fields are decomposed in the same way. Since the vector field is real, we have $\overline{A}_{\mu 3}=A_{\mu,3}$ and $\overline{A}_{\mu\pm}=A_{\mu\mp}$. We choose a basis such that the flux background is encoded only in the $\phi_3$ component,
\begin{equation}
	<\phi_{\pm}>=0 \qquad <\phi_3> = \frac{f}{2\sqrt{2}}(x_5-ix_6) .
	\label{eq:flux_background_nonabelian}
\end{equation}

We now compute the mass spectrum of the scalar fields. The Lagrangian contains the following scalar terms
\begin{align}
	\begin{split}
		\mathscr{L}_6 &\supset -\frac{1}{4}\left(\partial\overline{\phi}_3 + \overline{\partial}\phi_3+ \sqrt{2}g(|\phi_+|^2-|\phi_-|^2)^2 \right)^2 \\
		&-\frac{1}{2}\left(\partial\overline{\phi}_+ + \overline{\partial}\phi_- -\sqrt{2}g(\phi_3\overline{\phi}_+-\overline{\phi}_3\phi_-)\right)\left(\partial\overline{\phi}_- + \overline{\partial}\phi_+ +\sqrt{2}g(\phi_3\overline{\phi}_--\overline{\phi}_3\phi_+)\right)\\
		&= -\frac{1}{4}\left(\sqrt{2}f + \partial\overline{\varphi}_3 + \overline{\partial}\varphi_3+ \sqrt{2}g(|\phi_+|^2-|\phi_-|^2)^2 \right)^2\\
		&-\frac{1}{2}\left((-i\sqrt{2qf}a_-^{\dagger}-\sqrt{2}g\varphi_3)\overline{\phi}_+ + (-\sqrt{2qf}a_- +\sqrt{2}g\overline{\phi}_3)\phi_-\right)\\
		&\times \left((-i\sqrt{2qf}a_+^{\dagger}-\sqrt{2}g\overline{\varphi}_3){\phi}_+ + (-\sqrt{2qf}a_+ +\sqrt{2}g{\phi}_3)\overline{\phi}_-\right)\\
		&\supset -gf\left( |\phi_+|^2-|\phi_-|^2 \right) +gf\left(a_-^{\dagger}\overline{\phi}_+ + a_-\phi_-\right)\left(a_+^{\dagger}\phi_++a_+\overline{\phi}_-\right).
		\label{eq:gauge_scalar_mass_terms}
	\end{split}
\end{align}
Integrating over $T^2$ and using equations (\ref{eq:ladder_operators_non_susy}-\ref{eq:mode_functions_definition}), the four-dimensional effective Lagrangian reads
\begin{align}
	\begin{split}
		&\mathscr{L}_4^{mass} = -gf \sum_{n,j} \left( |\phi_{+;n,j}|^2-|\phi_{-;n,j}|^2 \right) \\
		&-gf\sum_{n,j;n',j'} \int dz^2 \left( \sqrt{n+1} \overline{\phi}_{+;n,j}\overline{\xi}_{n+1,j} - \sqrt{n}\phi_{-;n,j}\overline{\xi}_{n-1,j} \right)\\
		&\times\left(  \sqrt{n'+1}\phi_{+;n',j'}\xi_{n'+1,j'} - \sqrt{n'}\overline{\phi}_{-;n',j'}\xi_{n'-1,j'} \right).
	\end{split}
\end{align}
Using the orthonormality relation (\ref{eq:orthonormality_relation}), we find
\begin{align}
\begin{split}
	&\mathscr{L}_4^{mass} = gf|\phi_{-;0,j}|^2\\
    &- gf \sum_{n,j}\begin{pmatrix}\overline{\phi}_{+;n,j} & \phi_{-;n+2,j} \end{pmatrix}\begin{pmatrix} n+2 & -\sqrt{(n+1)(n+2)} \\ -\sqrt{(n+1)(n+2)} & n+1  \end{pmatrix} \begin{pmatrix} \phi_{+;n,j} \\\overline{\phi}_{-;n+2,j} \end{pmatrix}.
	\label{eq:mass_matrix_gauge}
\end{split}
\end{align}
The scalar mass spectrum is composed of:
\begin{enumerate}
	\item A tachyon $\phi_{-;0,j}$ with
	\begin{equation}
		m_{\phi^-_{0,j}}^2 = -gf.
		\label{eq:tachyon}
	\end{equation}
	This confirms the result of \cite{Buchmuller_2017} and equation \eqref{Bachas_equation}.
	\item A massless scalar$\phi_{-;1,j}$.
	\item One of the eigenstates of the scalar mass matrix is the tower of Nambu-Goldstone bosons
	\begin{equation}
		\phi_{n,j}^G = \sqrt{\frac{n+1}{2n+3}}\phi_{+;n,j} + \sqrt{\frac{n+2}{2n+3}}\overline{\phi}_{-;n+2,j}.
	\end{equation}
	\item The other eigenstate corresponds to the following massive scalars
	\begin{equation}
		\phi_{n,j}^M = - \sqrt{\frac{n+2}{2n+3}}\phi_{+;n,j} + \sqrt{\frac{n+1}{2n+3}}\overline{\phi}_{-;n+2,j}.
	\end{equation}
	Their masses are
	\begin{equation}
		m_{\phi^M_{n,j}}^2 = gf(2n+3).
	\end{equation}
\end{enumerate}

\subsection{Removing the tachyons}

As an alternative to tachyon condensation, the tachyons may be removed by shifting the whole mass spectrum. As a toy model, we introduce a scalar boson $\Phi$, with non-zero vacuum expectation value, in the adjoint representation of the gauge group $SU(2)$. We expect this vev to contribute to the masses of all gauge scalar fields and to remove the tachyons for sufficiently large values of the vev. We define
\begin{equation}
	\Phi = \Phi_3 T_3 + \Phi_- \frac{T_+}{\sqrt{2}} + \Phi_+\frac{T_-}{\sqrt{2}}.
\end{equation}
The Lagrangian describing the scalar boson and the gauge fields is
\begin{equation}
	\mathscr{L}_6 = -\frac{1}{2}\Tr F_{MN}F^{MN} - (D_M\Phi)^{\dagger}D^M\Phi - V(\Phi),
\end{equation}
where $V(\Phi)$ is the scalar potential
\begin{equation}
	V(\Phi) = - m^2\Phi^{\dagger}\Phi + \lambda (\Phi^{\dagger}\Phi)^2,
\end{equation}
and $D_M$ is the covariant derivative
\begin{equation}
	D_M \Phi = \partial_M - ig[A_M,\Phi].
\end{equation}
The minimum of this potential is not zero and yelds a finite vacuum expectation value for the scalar boson. By rotating the basis, we find
\begin{equation}
	<\Phi_3> = v = \sqrt{\frac{m^2}{2\lambda}}.
\end{equation}
In its matrix form, $\Phi$ reads
\begin{equation}
	\Phi = \frac{1}{2}\begin{pmatrix} \Phi_3 & \sqrt{2}\Phi_- \\\sqrt{2}\Phi_+ & -\Phi_3 \end{pmatrix}.
\end{equation}
The adjoint representation of $SU(2)$ is real, so that $\Phi^{\dagger}=\Phi$ and
\begin{equation}
	\Tr\Phi\Phi^{\dagger} = \frac{1}{2}\big( \Phi_3^2+2\Phi_+\Phi_- \big).
\end{equation}
Moreover,
\begin{equation}
	\Vec{\Phi}\cdot\Vec{\Phi} = \Phi_1^2+\Phi_2^2+\Phi_3^2 = \Phi_3^2+2\Phi_+\Phi_- = 2\Tr\Phi\Phi^{\dagger},
\end{equation}
such that
\begin{equation}
	\mathscr{L}_6 = -\frac{1}{2}\Tr F_{MN}F^{MN} - 2\Tr D_M\Phi D^M\Phi, \label{eq:Lag_gauge_and_scalar}
\end{equation}
with $<\Phi_3>=v\neq 0$. To compute the mass spectrum of the model, we write the covariant derivative of $\Phi$ in components:
\begin{align}
	\begin{split}
		D_M \Phi = &\left(\partial_M\Phi_3 -ig(A_M^-\Phi_+ - A_M^+\Phi_-) \right) T_3\\
  &+ \left( \partial_M\Phi_- -ig(A_M^3\Phi_- -A_M^-\Phi_3) \right) \frac{T_+}{\sqrt{2}} \\
		&+\left( \partial_M\Phi_+ -ig( A_M^+\Phi_3 - A_M^3\Phi_+ )\right) \frac{T_-}{\sqrt{2}}.
	\end{split}
\end{align}
Keeping only the scalar mass terms of the Lagrangian, we are left with
\begin{align}
	\begin{split}
		\mathscr{L}_6^{mass} &= 2gf \left( a_-^{\dagger}\Phi_-a_+^{\dagger}\Phi_+ + a_-\Phi_-a_+\Phi_+ \right) \\
  &+2i\sqrt{gf}gv \left( a_-^{\dagger}\Phi_-\overline{\phi}_- - a_-\Phi_-\phi_+ + \phi_-a_+^{\dagger}\Phi_+ - \overline{\phi}_+a_+\Phi_+ \right) \\
		& -(2g^2v^2 + gf)|\phi_+|^2 -(2g^2v^2-gf)|\phi_-|^2 \\
  &+gf\left(a_-^{\dagger}\overline{\phi}_+ + a_-\phi_-\right)\left(a_+^{\dagger}\phi_++a_+\overline{\phi}_-\right),
		\label{eq:gauge+scalar_mass_Lag}
	\end{split}
\end{align}
where we have used the decomposition
\begin{align}
	\begin{split}
		A_6^+ = \frac{1}{\sqrt{2}}(\phi_++\overline{\phi}_-) ,&\qquad A_5^+ = \frac{1}{\sqrt{2}i}(\phi_+-\overline{\phi}_-),\\
		A_6^- = \frac{1}{\sqrt{2}i}(\overline{\phi}_++{\phi}_-) ,&\qquad A_5^- = \frac{1}{\sqrt{2}i}(\phi_--\overline{\phi}_+).
	\end{split}
\end{align}
We now decompose the charged scalar fields $\Phi_{\pm}$ with respects to the Landau mode functions (\ref{eq:mode_functions_definition}), such that
\begin{align}
	\begin{split}
		\Phi_+ &= \sum_{n,j} \Phi_{+;n,j} \xi_{n,j},\\
		\Phi_- &= \sum_{n,j} \Phi_{-;n,j} \overline{\xi}_{n,j}.
	\end{split}
\end{align}
Inserting these definitions and (\ref{eq:mode_functions_definition}) into the Lagrangian (\ref{eq:gauge+scalar_mass_Lag}), and after using the orthonormality relation  (\ref{eq:orthonormality_relation}) as well as some translations in the Landau level, we find
\begin{align}
	\begin{split}
		\mathscr{L}&_4^{mass} = - 2gf \sum_{n,j}\left((2n+3)\Phi_{-;n+1,j}\Phi_{+;n+1,j} + \Phi_{-;0,j}\Phi_{+;0,j}\right) \\&+2\sqrt{gf}gv \sum_{n,j} \left( \Phi_{-;0,j}\overline{\phi}_{-;1,j} + \phi_{-;1,j}\Phi_{+;0,j}  +    \sqrt{n+2} \Phi_{-;n+1,j}\overline{\phi}_{-;n+2,j} \right. \\
  &+ \sqrt{n+1}\Phi_{-;n+1,j}\phi_{+;n,j} + \left. \sqrt{n+2}\phi_{-;n+2,j}\Phi_{+;n+1,j} +\sqrt{n+1} \overline{\phi}_{+;n,j}\Phi_{+;n+1,j} \right) \\
		& -(2g^2v^2 + gf)|\phi_{+;n,j}|^2 -(2g^2v^2-gf)(|\phi_{-;n+2,j}|^2+|\phi_{-;0,j}|^2+|\phi_{-;1,j}|^2) \\
		&-gf\sum_{n,j} \left( \sqrt{n+1} \overline{\phi}_{+;n,j} - \sqrt{n+2}\phi_{-;n+2,j}\right)\left(  \sqrt{n+1}\phi_{+;n,j} - \sqrt{n+2}\overline{\phi}_{-;n+2,j}\right) \\
        &-gf\sum_j|\phi_{-;0,j}|^2.
	\end{split}
\end{align}
Writing the Lagrangian in matrix form makes the computation of the mass spectrum and the expressions for the different eigenstates more explicit:
\begin{align}
	\begin{split}
		&\mathscr{L}_4^{mass} = - \sum_{n,j} \begin{pmatrix} \overline{\phi}_{+;n,j} & \phi_{-;n+2,j} & \Phi_{-;n+1,j} \end{pmatrix} \\
		&
		\times\begin{pmatrix} gf(n+2) +2g^2v^2 & -gf\sqrt{(n+1)(n+2)} & -2gv\sqrt{gf(n+1)}\\ -gf\sqrt{(n+1)(n+2)} & gf(n+1) +2g^2v^2 & -2gv\sqrt{gf(n+2)} \\
			-2gv\sqrt{gf(n+1)} &  -2gv\sqrt{gf(n+2)} & 2gf(2n+3)
		\end{pmatrix}\\
  &
		\times\begin{pmatrix} \phi_{+;n,j} \\ \overline{\phi}_{-;n+2,j} \\ \Phi_{+;n+1,j} \end{pmatrix}-2\sum_j \begin{pmatrix} \phi_{-;1,j} & \Phi_{-;0,j} \end{pmatrix}
		\begin{pmatrix} g^2v^2 & -gv\sqrt{gf} \\ -gv\sqrt{gf} & gf \end{pmatrix}\\
  &\times\begin{pmatrix}\overline{\phi}_{-;1,j} \\ \Phi_{+;0,j} \end{pmatrix} - (2g^2v^2-gf)|\phi_{-;0,j}|^2.
	\end{split}
\end{align}
The complete scalar mass spectrum is:
\begin{enumerate}
	\item The $3\times3$ matrix and the $2\times2$ matrices both have a massless eigenstate:
	\begin{align}
		\begin{split}
			\Phi^{G,1}_{n,j} &= \sqrt{\frac{n+1}{2n+3+v/f}}\phi_{+;n,j} + \sqrt{\frac{n+2}{2n+3+v/f}}\overline{\phi}_{-;n+2,j} \\
   &+ \sqrt{\frac{v/f}{2n+3+v/f}} \Phi_{+;n+1,j},\\
			\Phi^{G,2}_{j} &= \sqrt{\frac{1}{1+gv^2/f}}\overline{
				\phi}_{-;1,j} +\sqrt{\frac{1}{1+f/gv^2}}\Phi_{+;0,j}.
		\end{split}
	\end{align}
	\item The $3\times3$ matrix has two other eigenstates, which are massive, and the $2\times2$ matrix has a second eigenstate, which is also massive:
	\begin{align}
		\begin{split}
			\Phi^{M,1}_{n,j} &= - \sqrt{\frac{v}{f}\frac{n+1}{(2n+3+v/f)(2n+3)}}\phi_{+;n,j} \\
			&- \sqrt{\frac{v}{f}\frac{n+2}{(2n+3+v/f)(2n+3)}}\overline{\phi}_{-;n+2,j}\\
   &+ \sqrt{\frac{2n+3}{2n+3+v/f}} \Phi_{+;n+1,j},\\
			\Phi^{M,2}_{n,j} &=- \sqrt{\frac{n+2}{2n+3}}\phi_{+;n,j} + \sqrt{\frac{n+1}{2n+3}}\overline{\phi}_{-;n+2,j},\\
			\Phi^{M,3}_{j} &= -\sqrt{\frac{1}{1+f/gv^2}}\overline{\phi}_{-;1,j} + \sqrt{\frac{1}{1+gv^2/f}}\Phi_{+;0,j},
		\end{split}
	\end{align}
	and their masses are
	\begin{align}
		\begin{split}
			(m_{\Phi^{M,1}_{n,j}})^2 &= 2(gf(2n+3)+g^2v^2),\\
			(m_{\Phi^{M,2}_{n,j}})^2 &= gf(2n+3)+2g^2v^2,\\
			(m_{\Phi^{M,3}_{j}})^2 &= 2(gf+g^2v^2).
			\label{eq:scalar_masses_SU2}
		\end{split}
	\end{align}
	\item The last scalar would be a tachyon without the addition of the scalar $\Phi$. The fields $\phi_{-;0,j}$ are not tachyonic if
	\begin{equation}
		2gv^2 \geq f.
		\label{eq:inequation_tachyon}
	\end{equation}
\end{enumerate}
We have shown that by adding a scalar in the adjoint representation, the tachyons can be removed. Furthermore, for sufficiently large values of the vev of this scalar, the non-abelian flux background is perturbatively stable. The removal of tachyons depends on the physical parameters of the model. For fixed magnetic flux and vev, the presence of a tachyon depends on the gauge coupling. In terms of the coupling constants, the condition reads
\begin{equation}
    m^2\frac{g}{\lambda}\geq f.
\end{equation} In particular, it would be interesting to understand what happens when one renormalizes the theory to an energy scale below the limit $g=\frac{f}{2v^2}$. Note that the Lagrangian terms for $\Phi$ do not give rise to a $\phi_3$ mass term.

\subsection{Chiral fermions in the $SU(2)$ model without tachyon}

As a sanity check, we verify the chirality of fermions, say a $SU(2)$ doublet, in this tachyon-free model. The covariant derivative for this doublet is
\begin{equation}
	D_M = \partial_M + igqA_M.
\end{equation}
The complete Lagrangian must contain a covariant derivative term for the gauge field, the fermion doublet and the scalar triplet, a scalar potential, and an interaction term between the scalar field and the fermion doublet (Yukawa term). These terms read
\begin{align}
\begin{split}
	\mathscr{L}_6 &= -\frac{1}{2}\Tr F_{MN}F^{MN} + i \overline{\Psi}\Gamma^M\left( \partial_M + igq T^aA^a_M \right)\Psi\\
 &- (D_M\Phi)^{\dagger}D^M\Phi - V(\Phi) + \overline{\Psi}T^a\Psi\Phi^a,
\end{split}
\end{align}
where
\begin{equation}
	\Psi = \begin{pmatrix} \Psi^1 \\ \Psi^2 \end{pmatrix} \qquad \Psi^i=\begin{pmatrix} \psi^i \\ 0 \\ 0 \\ \overline{\chi}^i \end{pmatrix},
\end{equation}
and
\begin{equation}
	A_M = T^aA^a_M = \frac{1}{2}\begin{pmatrix} A_M^3 & \sqrt{2}A_M^- \\ \sqrt{2}A_M^+ & -A_M^3 \end{pmatrix}.
	\label{eq:vector_matrix_form}
\end{equation}
We assign charges $q_1$ and $q_2$ to the fermions of the doublet. We have already shown in Section \ref{sec:nonsusy_model} that the fermionic part of the Lagrangian (\ref{eq:non_susy_fermionic_action}) combines classical four-dimensional fermionic terms with the following internal component terms
\begin{equation}
	\mathscr{L}_{6f} \supset -\chi(\partial+\sqrt{2}gq\phi)\psi - \overline{\chi}(\overline{\partial}+\sqrt{2}gq\overline{\phi})\overline{\psi},
\end{equation}
where $\phi =  T^a \phi^a$. Using the matrix form \eqref{eq:vector_matrix_form}, one finds
\begin{align}
\begin{split}
	\mathscr{L}_{6f} &\supset - \begin{pmatrix} \chi^1 & \chi^2 \end{pmatrix}\begin{pmatrix} \partial + \sqrt{2}gq\phi_3 & 2gq\phi_i \\ 2gq\phi_+ & \partial-\sqrt{2}gq\phi_3 \end{pmatrix}\begin{pmatrix} \psi^1 \\ \psi^2 \end{pmatrix}\\
 &- \begin{pmatrix} \overline{\chi}^1 & \overline{\chi}^2 \end{pmatrix}\begin{pmatrix} \overline{\partial} + \sqrt{2}gq\overline{\phi}_3 & 2gq\overline{\phi}_+ \\ 2gq\overline{\phi}_- & \overline{\partial}-\sqrt{2}gq\overline{\phi}_3 \end{pmatrix}\begin{pmatrix} \overline{\psi}^1 \\ \overline{\psi}^2 \end{pmatrix}.
\end{split}
\end{align}
This Lagrangian contains interaction terms between the charged fields and mass terms for the fermions. Using annihilation and creation operators, we find the mass-squared operators for fermions. For $q_1=-q_2=1/2$, one has
\begin{align}
	\begin{split}
		\mathcal{M}^2_{\psi^1} &= gfa_+^{\dagger}a_+, \\
		\mathcal{M}^2_{\psi^2} &= gf(a_-^{\dagger}a_-+1), \\
		\mathcal{M}^2_{\chi^1} &= gfa_+^{\dagger}a_+ ,  \\
		\mathcal{M}^2_{\chi^2} &= gf(a_-^{\dagger}a_-+1).
		\label{fermions_mass_SU2}
	\end{split}
\end{align}
Finally, we add the Yukawa terms, with Yukawa coupling constant $h$:
\begin{align}
	\begin{split}
		\mathscr{L}_{Yuk} &= h\overline{\Psi}\Phi\Psi = \frac{1}{2}\begin{pmatrix} \overline{\Psi}^1 & \overline{\Psi}^2 \end{pmatrix}\begin{pmatrix} \Phi_3 & \sqrt{2}\Phi_- \\ \sqrt{2}\Phi_+ & -\Phi_3\end{pmatrix}\begin{pmatrix}\Psi^1 \\ \Psi^2\end{pmatrix}  = 0,
	\end{split}
\end{align}
where we have used the property that, for chiral six-dimensional fermions, the Lorentz invariant $\overline{\Psi}\Psi$ vanishes. Hence, there are no Yukawa terms in our simple model containing only one doublet. 

As a final sanity check, we compute the spectrum to check that it satisfies equation \eqref{Bachas_equation}. We find
\begin{align}
	\begin{split}
		&\mathscr{L}_{6}^{fermion-mass} = -\chi^1(\partial+gqf\overline{z})\psi^1 - \chi^2(\partial + gqf\overline{z})\psi^2 \\
  &- \overline{\chi}^1(\overline{\partial}+gfz)\overline{\psi}^1 - \overline{\chi}^2(\overline{\partial} +gqfz)\overline{\psi}^2\\
		&= i \sqrt{2gqf}\sum_{n,j;n',j'}\Big[\left( \chi^1_{n,j}a_+\psi^1_{n',j'} + \overline{\chi}_{n,j}^2a_-\overline{\psi}_{n',j'}^2 \right)\overline{\xi}_{n,j}\xi_{n',j'}\\
  &+ \left(  \chi^2_{n,j}a_+\psi_{n',j'}^2 + \overline{\chi}_{n,j}^1a_-\overline{\psi}_{n',j'}^1 \right)\xi_{n,j}\overline{\xi}_{n',j'}\Big]\\
		&= -\sqrt{2gqf(n+1)}\sum_{n,j}\Big[ \chi^1_{n,j}\psi^1_{n+1,j} + \overline{\chi}^2_{n,j}\overline{\psi}^2_{n+1,j} \\
  &+ \chi^2_{n,j}\psi^2_{n+1,j} + \overline{\chi}^1_{n,j}\overline{\psi}^1_{n+1,j} \Big].
	\end{split}
\end{align}
This confirms equation \eqref{Bachas_equation} in the case of fermions, which predicted $\delta\mathcal{M}^2_{fermions} = gf(n + 1/2 \pm 1/2)$. Indeed, the Lagrangian above implies 
\begin{align}
	\begin{split}
		m^2_{\psi_{n,j}} &= gfn,\\
		m^2_{\chi_{n,j}} &= gf(n+1).
	\end{split}
\end{align}
The left-handed zero modes are massless while the right-handed zero modes are massive.

\section{Conclusion}

In this paper, we have elaborated on the work of \cite{Buchmuller_2017,Buchmuller_2018}, exploiting the gauge vector field in a six-dimensional model with flux compactification on a torus to develop a toy model for the Higgs field.

The extra components of gauge fields, in the presence of a background magnetic flux, form a complex scalar field that remains massless. Unlike the case without flux, where quantum corrections introduce a dependence on the torus volume, the WL scalar is protected from one-loop corrections in $U(1)$ and $SU(2)$ gauge theories. However, in the $SU(2)$ case, the model contains a finite number of tachyonic states, which motivates further investigation into the true ground state, either through tachyon condensation or by removing the tachyons. The WL scalar is expected to remain massless at all-loop orders due to the model’s shift symmetry, thereby identifying the scalar as the Nambu-Goldstone boson of the translational symmetry in the internal space.

To move towards a more realistic Higgs model, we explored how the WL could acquire a non-vanishing mass while being protected from large quantum corrections. By adding an additional gauge symmetry, we exploited the principle that only one Nambu-Goldstone boson can arise from symmetry breaking. The two gauge fields introduce two WLs with shift symmetries that leave the Yukawa terms invariant, though only one combination of these symmetries preserves the remaining interaction terms. When the effective fluxes experienced by fermions are non-zero, both WLs behave as pseudo-Nambu-Goldstone bosons, acquiring non-vanishing masses at two-loop order. This raises the possibility of interpreting these scalars as Higgs-like bosons with finite masses that are protected from large quantum corrections. However, the WL scalars in this model are uncharged under the gauge symmetry, while the Standard Model Higgs is a charged $SU(2)$ doublet. A viable scenario would involve a WL associated with a $U(1)$ gauge field that transforms as a doublet under $SU(2)$, while still preserving the $U(1)$ shift symmetry.

In the second part of our study, we examined how to suppress tachyons in the $SU(2)$ gauge symmetry model. After confirming their presence, we introduced a scalar field $\Phi$ in the adjoint representation of $SU(2)$ and computed the full spectrum of charged scalar masses. This extension yields two charged Nambu-Goldstone bosons, which are absorbed via the Stückelberg mechanism by the charged vector fields $A^{\mu}_{\pm;n,j}$, along with three massive scalar modes and a potential tachyonic mode. We demonstrated that this remaining tachyon can be removed if the vacuum expectation value (vev) of the neutral component of the $SU(2)$ scalar triplet exceeds $\sqrt{\frac{f}{2g}}$. Our procedure is therefore subject to an additional hierarchy problem, which we did not discuss further in this toy model. The validity of this condition should be examined in any phenomenologically more realistic realization of the mechanism we have presented. Finally, we verified that the scalar field $\Phi$ preserves the chirality of fermions.

This work opens several avenues for future research: \begin{enumerate} \item Integrating the $SU(2)$ symmetry of the Standard Model to identify the Higgs boson with the WL of $U(1)$ gauge fields in the presence of magnetic flux, transforming as a doublet under $SU(2)$. It would also be crucial to verify that the shift symmetry is preserved in this scenario. \item We have shown that in a $U(1) \times U(1)$ gauge theory, WLs can acquire a non-vanishing mass protected from one-loop corrections. Future work could refine this result by calculating the two-loop quantum corrections to assess whether a realistic Higgs mass in the TeV range could emerge, especially in the context of large extra dimensions, where $1/R$ would be of the order of $~10$ TeV, thereby protecting the Higgs from one- and two-loop corrections. \item Investigating the condition \eqref{eq:inequation_tachyon} and the potential appearance of tachyons as a function of the gauge coupling could yield further insights. \item While our focus has been on addressing the hierarchy problem, a more straightforward application of this work lies in inflationary models. Indeed, the inflaton, like the scalar field in this model, is not charged, unlike the Higgs boson. A recent example is the model suggested in \cite{Hirose:2021xbs}, which uses the massless WL as an inflaton.
\item Finally, we should note that we neglect gravitational effects in this work. In particular, the backreaction of the flux on the background and its effects on the present considerations remains to be investigated.
\end{enumerate}

\acknowledgments

I would like to thank Emilian Dudas for his guidance and supervision during this project.\\

\vspace{1em}

\noindent \textbf{\large Data availability statement}\\

\noindent There are no new data associated with this article.

\bibliographystyle{jhep}
\bibliography{biblio}

\end{document}